\documentclass[aps,prl,preprintnumbers,amsmath,amssymb,twocolumn,superscriptaddress,showpacs,floatfix]{revtex4}
\usepackage{bbm}
\usepackage{mathrsfs}
\usepackage{amsmath,amssymb,bm,comment}
\usepackage{hyperref}
\usepackage{graphicx}
\usepackage{dcolumn}
\usepackage{bm}

\begin{document}

\title{Relativistic Quantum Kinetic Theory for Massive Fermions and Spin Effects}

\author{Jian-Hua Gao}
\affiliation{Shandong Provincial Key Laboratory of Optical Astronomy and Solar-Terrestrial Environment,
Institute of Space Sciences, Shandong University, Weihai, Shandong 264209, China}

\author{Zuo-Tang Liang}
\affiliation{Institute of Frontier and Interdisciplinary Science,
Key Laboratory of Particle Physics and Particle Irradiation (MOE), Shandong University, Qingdao, Shandong 266237, China}




\begin{abstract}
We present the complete first order relativistic quantum kinetic theory with spin for massive fermions derived from the Wigner function formalism in a concise form
that shows explicitly how the 32 Wigner equations reduce to 4 independent transport equations.
We solve the modified on-shell conditions to obtain the general solution and present the corresponding transport equations in three different forms that are suitable for different purposes.
We demonstrate how different spin effects arise from the kinetic theory by calculating
the chiral separation effect with mass correction, the chiral anomaly from the axial current
and the quantum magnetic moment density induced by vorticity and magnetic field.
We also show how to generate the global polarization effect due to spin vorticity coupling.
The formalism presented may serve as a practical theoretical framework to study different spin effects in relativistic fermion systems
encountered in different areas such as heavy ion, astro-particle and condensed matter physics as well.
\end{abstract}

\maketitle

{\it  Introduction.} ---
Spin plays an essential and fascinating role to probe the underlying structure of theories in different areas of physics. 
The recent observation~\cite{STAR:2017ckg,Adam:2018ivw} by STAR collaboration
of the global polarization~\cite{Liang:2004ph,Liang:2004xn,Betz:2007kg,Becattini:2007sr,Gao:2007bc} of $\Lambda$
hyperon in non-central heavy ion collisions opens new directions in the study of hot and dense nuclear matter
and motivates particularly further theoretical efforts on the physics of the global polarization effect (GPE)
and vorticity~
\cite{Becattini:2016gvu,Becattini:2017gcx,Becattini:2018duy,Fang:2016vpj,Pang:2016igs,Florkowski:2017ruc,Florkowski:2017dyn,Florkowski:2018ahw,Sun:2017xhx,Hattori:2019lfp}.
What is quite extraordinary in heavy ion collisions is that,  spin can emerge as a series of macroscopic collective effects such as,
besides the GPE observed by STAR~\cite{STAR:2017ckg,Adam:2018ivw},
the chiral magnetic effect (CME)~\cite{Vilenkin:1980fu,Kharzeev:2007jp,Fukushima:2008xe},
the chiral vortical effect (CVE), the chiral separation effect (CSE)~\cite{Vilenkin:1979ui,Erdmenger:2008rm,Banerjee:2008th,Son:2009tf,Landsteiner:2011cp,Gao:2012ix} and so on.
This is quite different from other high energy reactions and fascinating in its own way.
Because the hot and dense system produced in heavy-ion collisions expands and cools down very fast,
the natural and promising theoretical framework to deal with these novel collective quantum effects is the relativistic quantum kinetic theory (RQKT).
In recent years there has been a considerable amount of works and significant progresses on the chiral kinetic theory (CKT),
i.e., RQKT for massless fermions~\cite{Stephanov:2012ki,Son:2012zy,Chen:2012ca,Manuel:2013zaa,Manuel:2014dza,Chen:2014cla,Chen:2015gta,Hidaka:2016yjf,Mueller:2017lzw,Huang:2018wdl,Gao:2018wmr,Liu:2018xip}.
With the running of the beam energy scan program at RHIC and especially the discovery of global polarization at  relatively lower energies~\cite{STAR:2017ckg,Adam:2018ivw},
it becomes indispensable to develop a consistent and practical framework of RQKT to be capable of
treating various spin effects mentioned above for massive fermions. 

The covariant Wigner function formalism is a powerful and systematic quantum
kinetic approach~\cite{Heinz:1983nx,Elze:1986qd,Vasak:1987um,Zhuang:1995pd,Blaizot:2001nr},
which is very successful to derive CKT and describe CME and CVE consistently.
However RQKT for the massive fermions is very different from CKT because,  in addition to the particle density,
the spin polarization vector (SPV) or the axial current will also become dynamical quantity and
the Wigner equations are entangled with each other in much more complicated ways so that
transport equations are also much more complicated than those in the chiral limit.
So far, the existing works~\cite{Heinz:1983nx,Elze:1986qd,Vasak:1987um,Zhuang:1995pd,Blaizot:2001nr,Mueller:2019gjj}  on RQKT for massive fermions
are restricted to leading order transport equations for SPV, which is enough to
deal with the vector current or energy momentum tensor to the first order in $\hbar$.
However, they are incapable of accounting for how the GPE can be produced from vorticity
and how the chiral anomaly can arise in kinetic theory for massive fermions in a self-contained manner.

The aim of this paper is to derive RQKT with spin for massive Fermions to the first order in $\hbar$ completely 
and study the disentanglement of Wigner equations.
We then present the general solution of the modified on-shell conditions at this order.
After that, we apply them to various spin effects in relativistic fermion systems to demonstrate how they are derived from RQKT consistently.

{\it  The complete RQKT to the first order.} ---
We start with the Lorentz-covariant and gauge-invariant Wigner function $W(x,p)$ defined as \cite{Vasak:1987um}
\begin{eqnarray}
\label{wigner}
W_{\alpha\beta}
=\int\frac{d^4 y}{(2\pi)^4}
e^{-ip\cdot y}\langle \bar\psi_\beta\left(x_+\right)U\left(x_+,x_-\right)
\psi_\alpha\left(x_-\right)\rangle, \
\end{eqnarray}
where $x_{\pm}=x\pm y/2$, 
$\alpha$ and $\beta$ are spinor indices and $U\left(x_+,x_-\right)$ is the gauge link. 
This matrix can  be expanded in terms of the 16 independent $\Gamma$-matrices,
\begin{eqnarray}
\label{decomposition}
W=\frac{1}{4}\left[\mathscr{F}+i\gamma^5 \mathscr{P}+\gamma^\mu \mathscr{V}_\mu +\gamma^5\gamma^\mu \mathscr{A}_\mu
+\frac{1}{2}\sigma^{\mu\nu} \mathscr{S}_{\mu\nu}\right],
\end{eqnarray}
where $\mathscr{F}$, $\mathscr{P}$, $\mathscr{V}_{\mu}$,
$\mathscr{A}_{\mu}$ and $\mathscr{S}_{\mu\nu}$ are all real functions of $x$ and $p$ representing the scalar, pseudo-scalar,
vector, axial-vector and antisymmetric tensor components, respectively.
These 16 Wigner functions satisfy 32 Wigner equations derived from the Dirac equation \cite{Vasak:1987um}.
In this paper,  we will restrict ourselves to the external or self-consistent mean field approximation.
It turns out that we can choose the scalar $\mathscr{F}$ measuring the mass density
and the axial vector $\mathscr{A}_{\mu}$ denoting SPV  \cite{BialynickiBirula:1991tx} as
the independent  fundamental components and sort the 32 Wigner equations as follows.

(i) Eleven of them provide the expressions of other components $\mathscr{P}$,  $\mathscr{V}_{\mu}$
and $\mathscr{S}_{\mu\nu}$  in terms of $\mathscr{F}$ and  $\mathscr{A}_{\mu}$,
\begin{eqnarray}
\label{P}
\mathscr{P} &=&-\frac{\hbar}{2m} \nabla^\mu \mathscr{A}_{\mu}, \\
\label{Vmu}
\mathscr{V}_{\mu} &=& \frac{1}{m}p_\mu \mathscr{F}
- \frac{\hbar}{2m^2}\epsilon_{\mu\nu\rho\sigma}\nabla^\nu p^\rho \mathscr{A}^{\sigma},\\
\label{Smunu}
\mathscr{S}_{\mu\nu} &=&-\frac{1}{m} \epsilon_{\mu\nu\rho\sigma}p^\rho \mathscr{A}^{\sigma}
+\frac{\hbar}{2m^2}\left( \nabla_\mu p_\nu  -  \nabla_\nu p_\mu \right)\mathscr{F},\
\end{eqnarray}
where $\nabla^\mu=\partial^\mu_x - F^{\mu\nu}\partial_\nu^p$, 
$\epsilon_{\mu\nu\rho\sigma}$ is the completely anti-symmetric tensor with the convention $\epsilon^{0123}=1$,
and the electric  charge $e$ is included in the gauge field tensor $F^{\mu\nu}$ for brevity.

(ii) Five of the remaining 21 Wigner equations provide the modified on-shell conditions  for $\mathscr{F}$ and  $\mathscr{A}_{\mu}$,
\begin{eqnarray}
\label{F-onshell}
\left(p^2-m^2\right)\mathscr{F} &=& -\frac{\hbar}{m} p^\mu \tilde F_{\mu\nu}\mathscr{A}^{\nu} , \\
\label{A-onshell}
\left(p^2-m^2\right) \mathscr{A}_{\mu} &=& - \frac{\hbar}{m} p^\nu \tilde F_{\mu\nu}\mathscr{F},
\end{eqnarray}
where  $\tilde F_{\mu\nu}=\epsilon_{\mu\nu\rho\sigma}F^{\rho\sigma}/2$ is the dual field strength tensor.

(iii) Another five lead to the transport equations for $\mathscr{F}$ and  $\mathscr{A}_{\mu}$ respectively,
\begin{eqnarray}
\label{F-transport}
p\cdot \nabla\mathscr{F}&=&\frac{\hbar}{2m}p^\mu ( \partial_\lambda^x \tilde F_{\mu\nu} ) \partial_p^\lambda \mathscr{A}^{\nu},\\
\label{A-transport}
p\cdot \nabla\mathscr{A}_{\mu} &=&F_{\mu\nu} \mathscr{A}^{\nu}+\frac{\hbar}{2m} p^\nu (\partial_\lambda^x \tilde F_{\mu\nu}) \partial_p^\lambda \mathscr{F}.
\end{eqnarray}

(iv) There is another equation providing a subsidiary condition for the SPV $\mathscr{A}_{\mu}$, i.e.,
\begin{eqnarray}
\label{A-condition}
p\cdot  \mathscr{A} &=& 0.
\end{eqnarray}
It removes one component of $\mathscr{A}_{\mu}$ from independent dynamical components
and reduces the number of independent components to 4.

(v) Finally, it is remarkable to show that, with Eqs.~(\ref{P}-\ref{A-condition}) given above fulfilled,  all the rest 10 of the 32 Wigner equations
are satisfied automatically and are not listed here.

The general solution of the modified on-shell conditions given by Eqs.~(\ref{F-onshell}) and (\ref{A-onshell}) is given by,
\begin{eqnarray}
\label{F-1-b}
\mathscr{F} &=&\delta\left(p^2-m^2\right) \mathcal{F}+\frac{\hbar}{m} \tilde F_{\mu\nu} p^\mu \mathcal{A}^{\nu}\delta'\left(p^2-m^2\right),~~~  \\
\label{A-1-b}
\mathscr{A}_{\mu}&=& \delta\left(p^2-m^2\right)  \mathcal{A}_{\mu}  +\frac{\hbar}{m} p^\nu \tilde F_{\mu\nu}\mathcal{F}\delta'\left(p^2-m^2\right),
\end{eqnarray}
where $\mathcal{F}$ and $\mathcal{A}_{\mu}$ have to be nonsingular at $p^2-m^2=0$ and are determined by transport equations.
We find that it is more convenient to take $\mathcal{F}$ and $\mathcal{A}_{\mu}$ as independent fundamental quantities
instead of  $\mathscr{F}$ and $\mathscr{A}_{\mu}$ in practise. 
This is particularly the case when transient electromagnetic (EM) field  or scattering process is considered.
In this case, we have no  EM field or interaction in the far past or far future.
As can be seen from Eqs.~(\ref{F-1-b}) and (\ref{A-1-b}), both $\mathcal{F}$ and $\mathcal{A}_{\mu}$
contribute to $\mathscr{F}$ and $\mathscr{A}_{\mu}$ during the interaction,
while in the far past or future, they contribute only to $\mathscr{F}$ and $\mathscr{A}_{\mu}$ separately.

By substituting Eqs.~(\ref{F-1-b}) and (\ref{A-1-b}) into Eqs.~(\ref{F-transport}) and (\ref{A-transport}),
we obtain transport equations for $\mathcal{F}$ and $\mathcal{A}_{\mu}$ in the Lorentz covariant form,
\begin{eqnarray}
\label{F-eq-4v}
& & p\cdot\nabla 
\left[\mathcal{F}\delta\left(p^2-m^2\right)
+\frac{\hbar}{m}\tilde  F_{\mu\nu} p^\mu\mathcal{A}^{\nu}\delta'\left(p^2-m^2\right)\right]\nonumber\\
&=&\frac{\hbar}{2m}(\partial_\lambda^x \tilde F_{\mu\nu})\partial^\lambda_p\left[p^\mu \mathcal{A}^{\nu}\delta\left(p^2-m^2\right)\right], \\
\label{A-eq-4v}
& & p\cdot\nabla
\left[ \mathcal{A}_{\mu}\delta\left(p^2-m^2\right)
+\frac{\hbar}{m}  p^\nu \tilde F_{\mu\nu}\mathcal{F}\delta'\left(p^2-m^2\right)  \right]\nonumber\\
&=&  F_{\mu\nu}\left[ \mathcal{A}^{\nu}\delta\left(p^2-m^2\right)
+\frac{\hbar}{m} p_\lambda \tilde F^{\nu\lambda}\mathcal{F}\delta'\left(p^2-m^2\right)\right]\nonumber\\
& &+\frac{\hbar}{2m} (\partial_\lambda^x \tilde F_{\mu\nu}) \partial^\lambda_p \left[p^\nu \mathcal{F}\delta\left(p^2-m^2\right)\right].
\end{eqnarray}
Due to singular $\delta$ function or initial value problem in covariant Wigner functions~\cite{Ochs:1998qj},
Eqs.~(\ref{F-eq-4v}) and (\ref{A-eq-4v}) are not suitable for numerical evaluations.
For this purpose, we carry out the integration over $p_0$ to get rid of the singular $\delta$-function.
Integrating $p_0$ from $0$ to $\infty$ picks up particle's contributions while $-\infty$ to $0$ for anti-particles.
For brevity, we present only results for particle's contributions in the following.
The anti-particle's contribution is similar.
By doing this, we obtain the integrated transport equations and the subsidiary condition as,
\begin{eqnarray}
\label{Eq-F-a}
&&p\cdot \nabla  \mathcal{F}
=-\frac{\hbar\, p^\mu  }{2m} \left[ \frac{\tilde F_{\mu\nu}\bar  p^\lambda \nabla_\lambda }{E_p^2}
- (\bar \partial^\lambda_x \tilde F_{\mu\nu}) \bar \partial_\lambda^p \right]\mathcal{A}^{\nu},\\
\label{Eq-A-a}
&&p\cdot  \nabla  \mathcal{A}_{\mu}
=F_{\mu\nu} \mathcal{A}^{\nu}
- \frac{\hbar\, p^\nu }{2m }\left[\frac{ \tilde F_{\mu\nu}  \bar p^\lambda  \nabla_\lambda}{E_p^2}
-  (\bar\partial^\lambda_x \tilde F_{\mu\nu}) \bar\partial_\lambda^p \right]\mathcal{F},\nonumber \\
&&p \cdot  \mathcal{A} =0,
\end{eqnarray}
where $p$ is now on the mass-shell, i.e., $p_0=E_p=\sqrt{{\bf p}^2+m^2}$,
the derivative w.r.t. $p_0$ that was contained in the operator $\nabla^\mu$ is removed,
$\bar p = p - (n\cdot p) n= (0,{\bf p})$ and $n=(1,{\bf 0})$.
To do numerical calculations, it is more convenient to rewrite them in the 3-dimensional form,
\begin{eqnarray}
\label{Eq-F-3v}
& &\left(\nabla_t +{\mathbf v}\cdot{\pmb{\nabla}}  \right)\mathcal{F}\nonumber\\
&=&-\frac{\hbar }{2m E_p}\left[({\bf B}+{\bf E}\times {\bf v})
(  {\bf v}\cdot {\pmb \nabla} + E_p  \overleftarrow{\pmb \nabla}_x\cdot {\pmb \nabla}_p )\right.\nonumber\\
& &\left.\hspace{1.3cm} -({\bf B}\cdot {\bf v})(  {\bf v}\cdot {\pmb \nabla} + E_p \overleftarrow{\pmb \nabla}_x\cdot {\pmb \nabla}_p){\bf v}\right]
\cdot\pmb{ \mathcal{A} },\\
\label{Eq-A-3v}
& &\left(\nabla_t +{\mathbf v}\cdot{\pmb{\nabla}}  \right){\pmb{\mathcal{A}}}
+{\bf E}({\bf v}\cdot {\pmb{\mathcal{A}}})- {\bf B}\times\pmb{\mathcal{A}} \nonumber\\
&=&- \frac{\hbar\, }{2m E_p }({\bf B}+{\bf E}\times {\bf v})
(  {\bf v}\cdot {\pmb \nabla} + E_p  \overleftarrow{\pmb \nabla}_x\cdot {\pmb \nabla}_p )
\mathcal{F}, \hspace{0.8cm}
\end{eqnarray}
where ${\bf v}={\bf p}/E_p$, $\nabla_t=\partial_t + {\bf E}\cdot {\pmb{\nabla}_p}$,
${\pmb{\nabla}}=\pmb{\nabla}_x + {\bf B}\times {\pmb{\nabla}_p}$ and the left arrow over $\pmb{\nabla}_x$ denotes that
it  acts only on the EM fields on the left. Here we have chosen space components $\pmb{\mathcal{A}}$ as independent variables and the time component is $\mathcal{A}^0={\bf v}\cdot\pmb{ \mathcal{A} }$, obtained from the subsidiary condition.

Now, we have successfully derived the quantum kinetic equations in three different forms:
the covariant form [Eqs.~(\ref{F-eq-4v}) and (\ref{A-eq-4v})],
the integrated form [Eqs.~(\ref{Eq-F-a}) and (\ref{Eq-A-a})]
and the 3-dimensional form [Eqs.~(\ref{Eq-F-3v}) and  (\ref{Eq-A-3v})].
They are equivalent to each other and can be used for different purposes.
We emphasize in particular that the kinetic equations presented above are complete up to the first order of $\hbar$.
They are different from those obtained before in literature such as
\cite{Heinz:1983nx,Elze:1986qd,Vasak:1987um,Zhuang:1995pd,Blaizot:2001nr}
where only zeroth order of $\mathscr{A}_{\mu}$ was considered.
For comparison, in the following, we first present this ``simplified version''
then use the complete version given above to discuss how different spin effects arise from RQKT.

{\it  The simplified version.} ---
If we are interested only in quantum effects of the vector current  
$j_\mu =\int d^4p \mathscr{V}_{\mu}$ or the energy momentum tensor $T_{\mu\nu} = \int d^4p\, p_\nu \mathscr{V}_\mu$,
we need to consider only the zeroth order of $\mathscr{A}_{\mu}$ to get the first order of $j_\mu$ and $T_{\mu\nu}$.
In this case, the transport equations for $\mathscr{A}_{\mu}$ reduce to those obtained before in e.g.
\cite{Heinz:1983nx,Elze:1986qd,Vasak:1987um,Zhuang:1995pd,Blaizot:2001nr}.
We now rewrite them in a rather intuitive way.
We define
\begin{eqnarray}
P s_\mu ={\mathcal{A}_\mu}/{\mathcal{F}},\ \ \textrm{with}\ \ p\cdot s=0 \ \ \textrm{and}\ \ s^2=-1, ~~~
\end{eqnarray}
where $P s_\mu$ together denotes  the polarization vector, $P$ is the magnitude and $s_\mu$ is the unit vector in that direction.
To the zeroth order of $\hbar$,  $P$ and $s_\mu$ are decoupled from each other and from $\mathcal{F}$ as well, i.e.,
\begin{eqnarray}
\label{Polarization}
&&p\cdot\nabla 
\left[P\delta\left(p^2-m^2\right)\right]=0 ,\\
\label{spinvector}
&&p\cdot\nabla 
\left[ s_{\mu}\delta\left(p^2-m^2\right)\right]= F_{\mu\nu} s^{\nu}\delta\left(p^2-m^2\right).~~~~~~~
\end{eqnarray}
To the first order in $\hbar$, the equation of motion for $\mathcal{F}$ can be rewritten as,
\begin{eqnarray}
\label{F}
&&p\cdot\nabla
\left[\mathcal{F}\delta\left(p^2-m^2-2E_p\Delta E\right) \right]\nonumber\\
&&=\frac{\hbar}{2m}
(\partial^x_\lambda\tilde F^{\rho\sigma})\partial_p^\lambda
\left[p_\rho s_{\sigma} P \mathcal{F}\delta\left(p^2-m^2-2E_p\Delta E\right)\right].~~~~~~
\end{eqnarray}
This indicates that all  quantum corrections have been absorbed into the effective energy shift due to the interaction between the EM field and spin,
\begin{eqnarray}
\Delta E =-
{\hbar} P \tilde F^{\rho\sigma} p_\rho s_{\sigma}/{2m E_p}.
\end{eqnarray}
We call it the effective energy because the polarization $P$ is an ensemble averaged quantity that is unity for the pure state.
We see clearly that, under such approximations, the system remains unpolarized if the initial condition is unpolarized.
None of the spin effects such as the chiral anomaly, GPE and CSE can be derived in a self-contained manner from RQKT at this level.

{\it CSE with mass correction.} ---
The axial current can be induced along the external magnetic field imposed on the charged fermion system, which is the so called CSE.
With the general solution of  $ \mathscr{A}_{\mu}$ in Eq.~(\ref{A-1-b}),
we immediately obtain CSE  with mass correction.
We take the global equilibrium solution under constant EM field
\begin{eqnarray}
&&\mathcal{A_\mu}=0,\\
\label{FD-distribution}
&& \mathcal{F}=\frac{m}{2\pi^3}
\left[\frac{\theta\left(u\cdot p\right)}{e^{{( u\cdot p-\mu)}/{T}}+1}
+\frac{\theta\left(-u\cdot p\right)}{e^{-{(u\cdot  p-\mu)}/{T}}+1}\right],~~~
\end{eqnarray}
where $u^\mu$ is the fluid velocity and has been identified as $n^\mu$ in this calculation.
Substituting them into Eq.~(\ref{A-1-b}) and integrating over $p$, we obtain,
\begin{eqnarray}
\label{CSE}
j_5^\mu &=& \int d^4 p \mathscr{A}^{\mu} =  \sigma B^\mu ,
\end{eqnarray}
where the transport coefficient $\sigma$ is given by
\begin{eqnarray}
 \sigma= \frac{\hbar}{2\pi^2} \int_0^\infty d p
 \left(n_+ - n_-\right) , \ \ n_{\pm}=\frac{1}{e^{(E_p\mp\mu)/T}+1}. \
\end{eqnarray}
This is exactly the same result as that obtained in \cite{Lin:2018aon}, where Kubo formula and axial anomaly equation were used.
At the chiral limit of $m=0$, we obtain the well-known result $\sigma|_{m=0}=\hbar\mu/2\pi^2$
and at  zero temperature, we obtain
$\sigma|_{T\to 0}=\hbar\mu\sqrt{1-m^2/\mu^2}/2\pi^2$.

{\it Chiral anomaly.} ---
Chiral anomaly is a fascinating  quantum field effect that is absent at the classical level.
A proper RQKT should be able to account for this effect.
In the Wigner function formalism, chiral anomaly can be derived  directly from Eq.~(\ref{P}).
It should be pointed out that 
Eq.~(\ref{P}) is valid up to the second order in $\hbar$.
The first order correction of $\mathscr{A}^\mu$ contributes to the second order correction of $\mathscr{P}$.
In fact, as we will see below, it is the second order contribution of Eq.~(\ref{P}) that leads to the chiral anomaly.
By inserting Eqs.~(\ref{F-1-b}) and (\ref{A-1-b}) into Eq.~(\ref{P}), we obtain
$j_5 = \int d^4 p \mathscr{P}$ as,
\begin{eqnarray}
\label{P-2-b-int0}
 j_5 = -\frac{\hbar}{ 2m}  \partial_\mu^x j_{5}^\mu
+\frac{\hbar^2 E\cdot B}{2m^2} \int d^4 p \partial^\lambda_p \left[ p_\lambda  \mathcal{F}\delta'(p^2-m^2)\right], \hspace{0.3cm}
\end{eqnarray}
where 
we have dropped the nonsingular term that vanishes after the integration.
We choose two specific solutions in the limiting cases:
the free vacuum solution $\mathcal{F}={m}/{2\pi^3}$ and $\mathcal{A}_\mu=0$,
and the global equilibrium solution with Fermi-Dirac distribution given by Eq.~(\ref{FD-distribution}) in chiral limit of $m=0$.
By carrying out the integration in Eq.~(\ref{P-2-b-int0}), we obtain exactly the same correct coefficient of chiral anomaly in both cases.

Although the results of chiral anomaly obtained in the above-mentioned two cases are exactly the same,
it is interesting to note that they originate from different momentum regions.
For the free vacuum solution, there is no suppression at large momentum and the finite contribution comes from the large momentum,
while for the chiral limit with Fermi-Dirac distribution, there is an exponential suppression at large momentum,
the finite contribution is from the infrared region.
This indicates again further studying on where chiral anomaly comes from in RQKT~\cite{Mueller:2017lzw,Gao:2018wmr,Mueller:2019gjj} is
important and necessary.

{\it Quantum magnetization effect.} ---
The spin magnetic moment density is determined by the space components of the anti-symmetric tensor $\mathscr{S}_{\mu\nu}$.
We take the global equilibrium solution given by Eq.~(\ref{FD-distribution}) with constant thermal vorticity tensor
$\Omega_{\mu\nu}=\partial_\mu^x \beta_\nu - \partial_\nu^x\beta_\mu$ where $\beta_\mu ={u_\mu}/{T}$.
We insert it into Eq.~(\ref{Smunu}) to get $\mathscr{S}_{\mu\nu}$.
It follows that the magnetic moment density or magnetization vector in the local co-moving frame is given by,
\begin{eqnarray}
\label{MVE}
M_\mu =\frac{1}{2}\epsilon_{\nu\mu\alpha\beta} u^\nu \int d^4 p \mathscr{S}^{\alpha\beta}
= \hbar \kappa B_\mu -{\hbar}\rho\,\omega_\mu/m ,
\end{eqnarray}
where  $B_\mu = \tilde F_{\mu\nu} u^\nu$ is the magnetic field in the fluid rest frame,
$\omega_\mu = \epsilon_{\mu\nu\alpha\beta} u^\nu \partial^\alpha_x u^\beta /2$ is the vorticity vector,
the magnetic susceptibility $\kappa={m}\int{d p} (n_+ + n_-)/{2\pi^2}E_p$ and $\rho$ is the charge density.

The presence of $\hbar$ on r.h.s. shows that magnetization due to magnetic field or vorticity is a pure quantum effect.
As we all know, charged vorticity can induce magnetic moment even in classical physics.
But the result  presented here is different from this classical mechanism and it represents ensemble average of the
intrinsic magnetic moment that does not depend on the origin of the coordinate that we choose.
The term provides the relationship between the vorticity and the induced spin magnetization.
We emphasize that, though obtained with the Fermi-Dirac distribution, the vorticity term in  Eq.~(\ref{MVE}) is
valid if $\mathcal{F}$ is chosen as any function of $p\cdot u$.

{\it GPE generation.} ---
Let us consider a fermionic system subject to a transient external  EM field.
We suppose that at $t=0$, there is no EM field, and the system is unpolarized so that ${\pmb{\mathcal{A}}}=0$ and $\mathcal{F}\neq 0$.
The EM field comes up at the next moment and the evolution of ${\pmb{\mathcal{A}}}$ after an infinitesimal time interval is determined by,
 \begin{eqnarray}
 \label{A-evolution}
\nabla_t {\pmb{\mathcal{A}}}
&=&- \frac{\hbar\, }{2m E_p }({\bf B}+{\bf E}\times {\bf v})
(  {\bf v}\cdot {\pmb \nabla} + E_p  \overleftarrow{\pmb \nabla}_x\cdot {\pmb \nabla}_p ) \mathcal{F}. \hspace{0.8cm}
\end{eqnarray}
We see clearly that the polarization is generated along the magnetic field due to the inhomogeneous distribution of $\mathcal{F}$ in phase space.

From Eq.~(\ref{A-evolution}), one might also conclude that, if there is no external EM field at $t=0$, the transport equations for $\mathcal{F}$
and ${\pmb{\mathcal{A}}}$ would decouple from each other so that GPE became impossible from unpolarized initial state.
This is not true in practice since the self-consistent EM field is determined
by Maxwell's equations
$\partial_\mu F^{\mu\nu}=  e^2 j^\nu$ 
($e^2$ is due to the extra $e$ in $F^{\mu\nu}$) that close with transport equations that in turn determine the current $j_\mu$.
The relationship between the vorticity  and EM field
can be seen more directly if we re-write Maxwell's equations as quadratic equations for the field tensor, i.e.~\cite{Elze:1986qd},
\begin{equation}
\partial_\lambda\partial^\lambda F_{\mu\nu}= e^2\left(\partial_\mu j_\nu - \partial_\nu j_\mu\right).
\end{equation}
We see explicitly that it is the vorticity of $j_\mu$ instead of the current itself that induces EM field tensor.
The induced EM field will couple $\pmb{\mathcal{A}}$ and $\mathcal{F}$ together and generate the polarization along the direction of vorticity.
This provides a mechanism to generate the GPE from vorticity in the current.
We emphasize that here we have no collision terms but only relativistic quantum effects.
Further studies can be carry out numerically for specified systems in future works.

{\it Summary and outlook.} ---
By choosing the scalar $\mathscr{F}$ and the axial vector $\mathscr{A}_{\mu}$ as independent components,
we have presented the RQKT for particle with spin-1/2 up to the first order in $\hbar$ derived from
the covariant Wigner function formalism completely and in a concise form.
This form shows explicitly that the 32 Wigner equations are in practice replaced by the 4 independent transport equations.
We have solved the modified on-shell conditions and obtained the general solution.
Using this we have presented the transport equations in three different forms, the covariant, integrated and 3-dimensional forms,
that can be used for different purposes.
We have demonstrated that different spin effects such as the chiral anomaly,
CSE, quantum magnetization effect and GPE can arise from the RQKT automatically.


We have pointed out in particular that these spin effects arise only when we consider Wigner equations up to the first order in $\hbar$ completely.
None of them arises automatically if we consider only the current and energy momentum tensor to the first order
but take the axial vector $\mathscr{A}_{\mu}$ only to the zeroth order.

By calculating the axial current $j_{5\mu}=\int d^4p\mathscr{A}_{\mu}$ and $j_5 = \int d^4 p \mathscr{P}$, we have successfully
shown that CSE with mass correction and the chiral anomaly are obtained automatically in the self-contained manner
 and lead to correct forms as obtained in other approaches.
%
%
We have also calculated the  quantum magnetization effect
and obtained the simple relationship $M_\mu =\hbar \kappa B_\mu -{\hbar}\rho\,\omega_\mu/m$.
We have also shown in particular that, although no collision term is considered yet,
GPE can be generated from current vorticity due to back reactions of EM fields from Maxwell's equations.
The formalism presented provides a practical theoretical framework in RQKT to study different spin effects
in different relativistic quantum systems.
It can be extended to higher orders in a straightforward manner
and is applicable not only in heavy-ion physics but also other areas such as astrophysics as well as condensed matter physics
where polarization effects are involved in massive fermion systems.

\textit{Acknowledgments.} ---
We thank Q. Wang for helpful discussions.
This work was supported in part by  the National Natural
Science Foundation of China  under 
Nos. 11890713 and 11675092, and
the Natural Science Foundation of Shandong Province under 
No. JQ201601.

\textit{Note added.} --- While this work was being completed, we became aware of
 the work \cite{Sheng:2019} which  also derives the RQKT from Wigner-function formalism.

\end{document}